\documentclass[aip, amsmath, amssymb, reprint]{revtex4-1}

\usepackage{graphicx}
\usepackage{dcolumn}
\usepackage{bm}
\usepackage{hyperref}
\hypersetup{breaklinks=true,colorlinks=true,linkcolor=blue,urlcolor=blue,citecolor=blue}
\usepackage{mathptmx}
\usepackage{etoolbox}
\usepackage{comment}

\makeatletter
\def\@email#1#2{
 \endgroup
 \patchcmd{\titleblock@produce}
  {\frontmatter@RRAPformat}
  {\frontmatter@RRAPformat{\produce@RRAP{*#1\href{mailto:#2}{#2}}}\frontmatter@RRAPformat}
  {}{}
}
\makeatother
\begin{document}

\title[On Ultrafast X-ray Methods for Magnetism]{On Ultrafast X-ray Methods for Magnetism}

\author{R. Plumley}
\affiliation{Stanford Institute for Materials and Energy Science, Stanford University and SLAC National Accelerator Laboratory, Menlo Park, California 94025}
\affiliation{Linac Coherent Light Source, SLAC National Accelerator Laboratory, Menlo Park, CA 94720}
\affiliation{Physics Dept., Carnegie Mellon University,
Pittsburgh, PA 15213}

\author{S. R. Chitturi}
\affiliation{Linac Coherent Light Source, SLAC National Accelerator Laboratory, Menlo Park, CA 94720}
\affiliation{Department of Materials Science and Engineering, Stanford University, Stanford, CA 94305}

\author{C. Peng}
\affiliation{Stanford Institute for Materials and Energy Science, Stanford University and SLAC National Accelerator Laboratory, Menlo Park, California 94025}

\author{T. A. Assefa}
\affiliation{Stanford Institute for Materials and Energy Science, Stanford University and SLAC National Accelerator Laboratory, Menlo Park, California 94025}
\affiliation{Division of Synchrotron Radiation Research, Department of Physics, Lund University, Solvegatan 14, 22100 Lund, Sweden}

\author{N. Burdet}
\affiliation{Stanford Institute for Materials and Energy Science, Stanford University and SLAC National Accelerator Laboratory, Menlo Park, California 94025}
\affiliation{Linac Coherent Light Source, SLAC National Accelerator Laboratory, Menlo Park, CA 94720}

\author{L. Shen}
\affiliation{Stanford Institute for Materials and Energy Science, Stanford University and SLAC National Accelerator Laboratory, Menlo Park, California 94025}
\affiliation{Division of Synchrotron Radiation Research, Department of Physics, Lund University, Solvegatan 14, 22100 Lund, Sweden}

\author{A. H. Reid}
\affiliation{Linac Coherent Light Source, SLAC National Accelerator Laboratory, Menlo Park, CA 94720}

\author{G. L. Dakovski}
\affiliation{Linac Coherent Light Source, SLAC National Accelerator Laboratory, Menlo Park, CA 94720}

\author{M. H. Seaberg}
\affiliation{Linac Coherent Light Source, SLAC National Accelerator Laboratory, Menlo Park, CA 94720}

\author{F. O'Dowd}
\affiliation{Linac Coherent Light Source, SLAC National Accelerator Laboratory, Menlo Park, CA 94720}

\author{S. A. Montoya}
\affiliation{Center for Memory and Recording Research, University of California–-San Diego, La Jolla, California 92093}

\author{H. Chen}
\affiliation{Physics Department, Northeastern University, Boston, MA 02115}
\affiliation{Stanford Institute for Materials and Energy Science, Stanford University and SLAC National Accelerator Laboratory, Menlo Park, California 94025}

\affiliation{Linac Coherent Light Source, SLAC National Accelerator Laboratory, Menlo Park, CA 94720}

\author{A. Okullo}
\affiliation{Department of Physics and Astronomy, Howard University, Washington DC, 20059}

\author{S. Mardanya}
\affiliation{Department of Physics and Astronomy, Howard University, Washington DC, 20059}

\author{S. D. Kevan}
\affiliation{Department of Physics, University of Oregon, Eugene, OR 97401}
\affiliation{Advanced Light Source, Lawrence Berkeley National Laboratory, Berkeley, CA 94720}

\author{P. Fischer}
\affiliation{Materials Sciences Division, Lawrence Berkeley National Laboratory, Berkeley, CA 94720}

\author{E. E. Fullerton}
\affiliation{Center for Memory and Recording Research, University of California–-San Diego, La Jolla, California 92093}
\affiliation{Department of Electrical and Computer Engineering, University of California–-San Diego, La Jolla, California 92093}

\author{S. K. Sinha}
\affiliation{Department of Physics, University of California–-San Diego, La Jolla, California 92093}

\author{W. Colocho}
\affiliation{Linac Coherent Light Source, SLAC National Accelerator Laboratory, Menlo Park, CA 94720}

\author{A. Lutman}
\affiliation{Linac Coherent Light Source, SLAC National Accelerator Laboratory, Menlo Park, CA 94720}
\author{F.-J. Decker}
\affiliation{Linac Coherent Light Source, SLAC National Accelerator Laboratory, Menlo Park, CA 94720}

\author{S. Roy}
\affiliation{Advanced Light Source, Lawrence Berkeley National Laboratory, Berkeley, CA 94720}

\author{J. Fujioka}
\affiliation{Department of Applied Physics and Quantum-Phase Electronics Center (QPEC), University of Tokyo, Hongo, Tokyo 113-8656, Japan}
\affiliation{Graduate School of Pure and Applied Science,
University of Tsukuba, Tsukuba, Ibaraki, 305-8573, Japan}

\author{Y. Tokura}
\affiliation{Department of Applied Physics and Quantum-Phase Electronics Center (QPEC), University of Tokyo, Hongo, Tokyo 113-8656, Japan}
\affiliation{RIKEN Center for Emergent Matter Science (CEMS), Wako 351-0198, Japan}

\author{M. P. Minitti}
\affiliation{Linac Coherent Light Source, SLAC National Accelerator Laboratory, Menlo Park, CA 94720}

\author{J. A. Johnson}
\affiliation{Swiss Light Source, Paul Scherrer Institut (PSI), 5232 Villigen PSI, Switzerland}

\author{M. Hoffmann}
\affiliation{Linac Coherent Light Source, SLAC National Accelerator Laboratory, Menlo Park, CA 94720}%

\author{M. E. Amoo}
\affiliation{Department of Mechanical Engineering, Howard University, Washington DC, 20059}

\author{A. Feiguin}
\affiliation{Physics Department, Northeastern University, Boston, MA 02115}

\author{C. Yoon}
\affiliation{Linac Coherent Light Source, SLAC National Accelerator Laboratory, Menlo Park, CA 94720}
\author{J. Thayer}
\affiliation{Linac Coherent Light Source, SLAC National Accelerator Laboratory, Menlo Park, CA 94720}

\author{Y. Nashed}
\affiliation{SLAC National Accelerator Laboratory, Menlo Park, California 94025}

\author{C. Jia}
\affiliation{Stanford Institute for Materials and Energy Science, Stanford University and SLAC National Accelerator Laboratory, Menlo Park, California 94025}

\author{A. Bansil}
\affiliation{Physics Department, Northeastern University, Boston, MA 02115}

\author{S. Chowdhury*}
\email{sugata.chowdhury@howard.edu}
\affiliation{Department of Physics and Astronomy, Howard University, Washington DC, 20059}

\author{A. M. Lindenberg*}
\email{aaronl@stanford.edu}
\affiliation{Stanford Institute for Materials and Energy Science, Stanford University and SLAC National Accelerator Laboratory, Menlo Park, California 94025}
\affiliation{Department of Materials Science and Engineering, Stanford University, Stanford, CA 94305}

\author{M. Dunne}
\affiliation{Linac Coherent Light Source, SLAC National Accelerator Laboratory, Menlo Park, CA 94720}

\author{E. Blackburn}
\affiliation{Division of Synchrotron Radiation Research, Department of Physics, Lund University, Solvegatan 14, 22100 Lund, Sweden}

\author{J. J. Turner*}
\email{joshuat@stanford.edu}
\affiliation{Stanford Institute for Materials and Energy Science, Stanford University and SLAC National Accelerator Laboratory, Menlo Park, California 94025}
\affiliation{Linac Coherent Light Source, SLAC National Accelerator Laboratory, Menlo Park, CA 94720}

\date{\today}

\begin{abstract}
With the introduction of x-ray free electron laser sources around the world, new scientific approaches for visualizing matter at fundamental length and time-scales have become possible. As it relates to magnetism and `magnetic-type' systems, advanced methods are being developed for studying ultrafast magnetic responses on the time-scales at which they occur. We describe three capabilities which have the potential to seed new directions in this area and present original results from each: pump-probe x-ray scattering with low energy excitation, x-ray photon fluctuation spectroscopy, and ultrafast diffuse x-ray scattering. By combining these experimental techniques with advanced modeling together with machine learning, we describe how the combination of these domains allows for a new understanding in the field of magnetism. Finally, we give an outlook for future areas of investigation and the newly developed instruments which will take us there.
\end{abstract}
\maketitle

\tableofcontents
\section{Introduction}
In recent years, remarkable new phases of matter have been both predicted and measured, such as quantum spin liquids, skyrmions, strongly spin-orbit coupled materials, quantum spin Hall insulators, and helical topological superconductors.  These phases often arise due to delicate combination of multiple interactions such as quantum confinement or magnetic frustration, and can display magnetic features with both short-range and long-range order ranging from sub-nanometer to micron length scales \cite{keimer2017physics}. Quantum materials exhibiting this behavior hold promise for highly efficient electronic- and spin-transport as well as tunability for technological applications~\cite{giustino20212021, cao2018unconventional}. One such field is that of spintronics, where the magnetic degree of freedom can be used as a knob for new functionalities and enable new abilities to control materials~\cite{vzutic2004spintronics, hirohata2020review}. 


A common theme in ``quantum engineering" of materials and devices is the ability to temporarily drive one phase into another, usually by the introduction of a symmetry-breaking mechanism or radiative excitation~\cite{du2021engineering, gong2019two, bogdanov-prl-2001, sie2019ultrafast, qian2014quantum, liu2014tuning, oka2009photovoltaic}, or to create new transient states of matter \cite{wandel-science-2022, tengdin2018critical, shan2021giant}.  This necessitates a robust theoretical understanding of how systems respond to stimulation at ultrafast time-scales. One such example in the field of magnetism is the 2D Van der Waals materials~\cite{joy1992magnetism, wildes2015magnetic}. Here, theoretical models can be tested directly, such as the Berezinskii-Kosterlitz-Thouless transition which predicts topological order. In the 2D limit, magnetic fluctuations can become dominant, and can also act to mediate the formation of other types of hidden quantum phases \cite{burch2018magnetism}. Probing such fluctuations requires the development of new experimental tools with sensitivity on the requisite time- and length- scales.




Another rapidly growing field of study is the understanding and control of discrete topological objects, such as magnetic skyrmions. These have been shown to respond to small magnetic fields with incredible implications for technological applications such as computer memory~\cite{finocchio2016magnetic, giustino20212021}.  While the motion of individual spin-moments can be described on the nanosecond time-scale by the Landau-Lifshitz-Gilbert equation, descriptions of spin-dynamics alone are insufficient for getting the full picture, since the emergence of skyrmion spin-textures arise from complicated competitions between a host of magnetic interactions, and can involve quasi-particles which are many lattice units in size. These structures exhibit dynamics that range across many time-scales from microseconds to the ultrafast, and can span length-scales much larger than the unit cells of their parent lattice~\cite{muhlbauer-2009-science, bogdanov1994thermodynamically, bogdanov-prl-2001,li-2021-nature}. 

Many powerful methods exist and have continued to be developed for understanding dynamics related to novel magnetic phenomena.  Nanoscale textures such as those seen in skyrmion crystals for instance are often studied using Lorentz Transmission Electron Microscopy (LTEM), a powerful tool for analyzing magnetic structure with down to 2~nm resolution~\cite{tang2019lorentz}.  However, LTEM is not sensitive to the rapid changes in the magnetic structure discussed above due to long integration times lasting seconds to minutes.  Despite challenges associated with the low readout-rates of LTEM detectors, efforts to push LTEM to the ultra-fast regime using stroboscopic methods are currently being pursued~\cite{moller2020few, da2018nanoscale}. Small-Angle Neutron Scattering (SANS) and Neutron Spin-Echo spectroscopy (NSE) are additional important methods for accessing the dynamic structure factor down to nanosecond time-scales with atomic precision, but can sometimes be challenging compared to x-ray sources due to the low absorption of neutrons in some materials, as compared to electromagnetic radiation~\cite{nakajima2020crystallization, kawecki2019direct}.  This aspect can hinder the capability of neutron methods for studying thin samples and systems undergoing rapid time-development, where neutron signal flux has to be averaged over long times.  In order to make continued progress, novel methods are needed to access magnetic structure and excitations at relevant length-scales as above, but that additionally can access dynamic phenomena at ultrafast timescales.

Complementary tools now are available which can target some of the modern topics in magnetism for the study of dynamics at the atomic scale using X-ray Free-Electron Lasers (X-FEL). This effort has a long history \cite{kirilyuk2010ultrafast}, from the first demonstrated rapid quenching of magnetic order \cite{beaurepaire1996ultrafast} to the early use case of using the Stanford Linear Accelerator to create short magnetic field pulses to read and write in magnetic recording media \cite{siegmann1995magnetism}, and now these new machines are creating enhanced capabilities in this area (For some good reviews, see \cite{malvestuto2018ultrafast,jeppson2021capturing,Durr2020}). These type of sources have had impact in many scientific research areas \cite{Bostedt-2016-RMP}, notably in wide-ranging areas such as atomic and molecular science~\cite{young2010femtosecond, gomez2014shapes, Glownia-2010-OptExp}, astrophysics \cite{Bernitt-2012-Nature,Vinko-2012-Nature}, condensed matter and materials physics\cite{Beaud-2014-NatMat,Kubacka-2014-Science,wandel-science-2022,Mankowsky-2014-Nature}, and structural biology \cite{Boutet-2012-Science, Chapman-2011-Nature}. This versatility is mainly owing to the ability to provide femtosecond x-ray pulse durations, large pulse energies, and repetition rates on the order of MHz~\cite{Bostedt-2016-RMP,white2015linac, altarelli2011european, schoenlein2017linac}. The photon energy is also typically tunable, allowing for element-specific resonant x-ray scattering for instance, which can isolate and enhance the magnetic structure signal from the sample~\cite{decker2022tunable, Caviglia-2013-PRB, Lee-2012-NatCom, Chuang-2013-PRL,Forst-2014-PRB, Foerst-2011-PRB, Foerst-2015-NatMat, shen-2020-prb}. These next generation light sources hold vast potential for driving the field of magnetism to new heights. 


It is in this context that we provide a forward-looking perspective in the field of magnetism focused on methods based on X-FEL radiation, especially as it relates to the ultrafast regime. We specifically focus on three methods which could be transformative in this area. With a close coupling between theory, combined with recent machine-assisted analysis techniques, we outline recent progress which has occurred by optimizing this overlap. After a description of the experimental methods, we report results on a cross-section of magnetic systems. We support this with theoretical tools which have potential for advanced time domain studies, as well as machine-learning developments focused on the single photon measurement. Finally, we provide an outlook for spin-sensitive studies in the ultrafast regime. We conclude with a discussion on the scientific prospects which will be available with the latest accelerator modes and new instruments being constructed at the LCLS-II.

\begin{figure*}
    \centering
    \includegraphics[width=.9\linewidth]{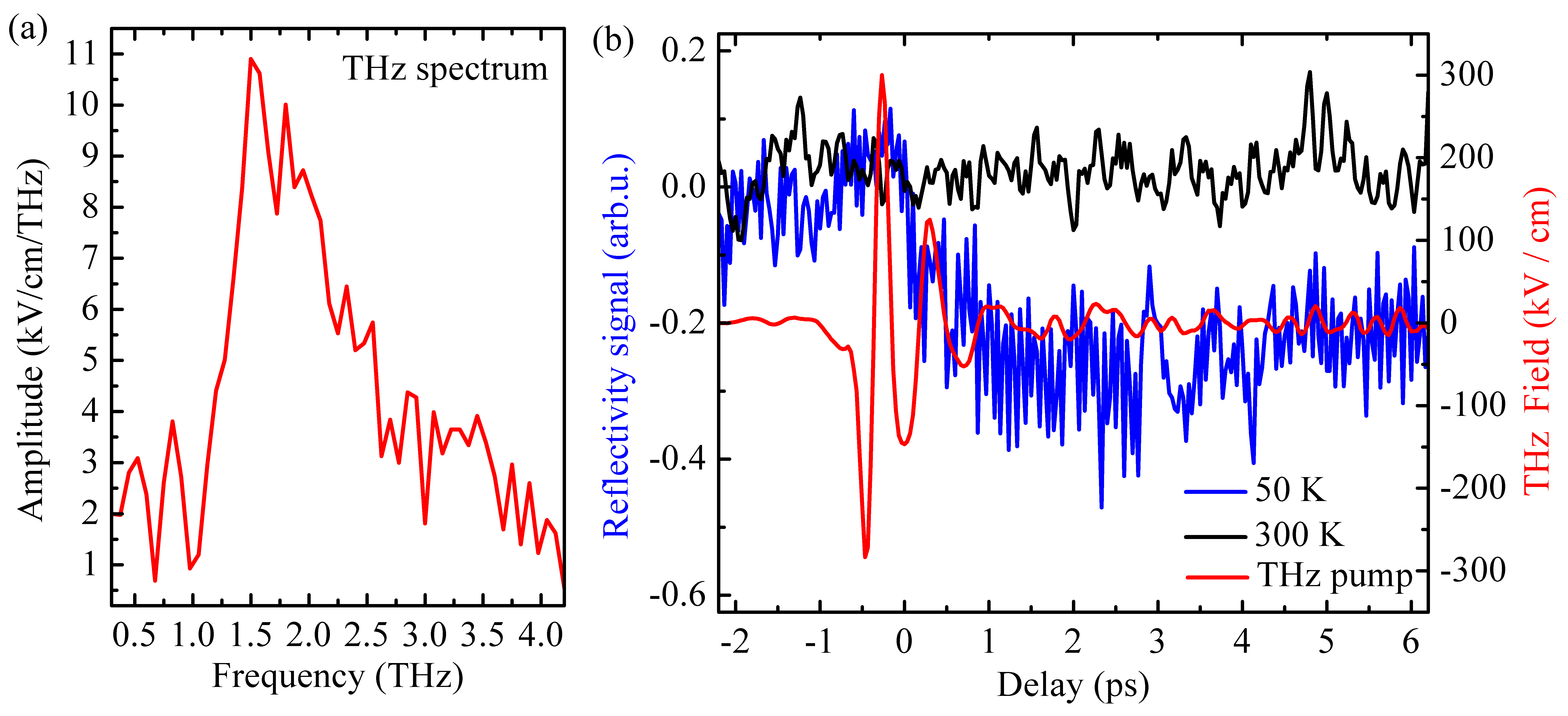}
    \caption{
Amplitude spectrum of the THz pump pulse obtained
from the Fourier transform of its time trace. (b) Temporal evolution
of the 800 nm optical reflectivity (left axis) at 50 K (blue) and 300 K (black)
after p-polarized THz photoexcitation of the manganite single crystal, Nd$_{1-x}$Sr$_{1+x}$MnO$_{4}$ ($\it{x}$\ =\ 2/3). The lower temperature is below the Neel temperature in this crystal and is likely demonstrating enhanced sensitivity to the magnetic order. The time trace of a single
THz pulse is shown in red (right axis).
    }
    \label{fig:thz}
\end{figure*}

\section{Ultrafast experimental methods}

\subsection{THz / X-ray pump-probe scattering}
\label{pumpprobesec}
One area of focus combines femtosecond x-ray probes with THz excitation of materials, especially in the area of magnetism \cite{Hoffmann-2012-SRN}. Using resonant x-ray scattering, the magnetic structure can be directly probed, such as the $L-$edge, responsible for $p$-to-$d$ transitions of the magnetic ion. A powerful use of X-FELs is to combine this sensitivity of the magnetic structure with THz excitation. With short-pulsed THz radiation, low energy modes can be directly excited for spin relaxation, spin enhancement, or coherent spin control \cite{Kubacka-2014-Science}. For instance, ultrashort THz pump pulses can directly couple to electromagnons \cite{Kubacka-2014-Science}. By using soft x-ray scattering at the Mn $L-$edge to probe the spin state in TbMnO$_3$, strong field THz was used to both excite and study the response of an electromagnon excitation \cite{Dakovski-2015-JSR, Turner-2015-JSR}. With current developments underway at the LCLS-II (see Sec.\ref{sec:enh-inst} and Sec.\ref{sec:fut-inst}), THz pumping will soon be possible while probing with high-repetition rate soft x-rays for direct sensitivity to different types of electronic ordering.  

Another example is in non-resonant THz pumping, where THz excitation has been shown to have spin sensitivity. Preliminary measurements were carried out  \cite{Schlotter-2012-RSI} on a manganite single crystal of Nd$_{1-x}$Sr$_{1+x}$MnO$_{4}$ ($\it{x}$\ =\ 2/3) consisting of non-resonant THz pumping the system, and x-ray or optical probe. This sample was grown by the floating-zone method \cite{kimura-2001-prb} and polished along the (110) direction. Complementary measurements of optical conductivity \cite{fujioka-2010-prb}, 800 nm pump-probe measurements with soft x-ray resonant probe \cite{shen-2020-prb}, and time-resolved optical reflectivity measurements \cite{shen-2020-prb} have all been carried out. A high-power Ti:sapphire-based laser (1.55 eV) with a 50-fs pulse duration and 120-Hz repetition rate which was split and cross-polarized for the pump pulses, which gives a temporal resolution of about 75 fs. THz generation was produced through non-linear rectification using a organic salt crystal (DAST) to generate field strengths in excess of 300 kV/cm (See Fig.\ref{fig:thz}a). By cooling the sample with a liquid He cryostat in UHV \cite{Doering-2011-RSI,Turner-2015-JSR}, the crystal could be studied below the Neel temperature of 90 K \cite{shen-2020-prb}.

The main THz results demonstrate how sensitive the magnetic ordering is to THz excitation, and are shown in Fig.\ref{fig:thz}b. This illustrates the reflectivity response to high-field, short-pulse THz radiation centered at around 1.5 THz, the frequency spectrum of which is shown in Fig.\ref{fig:thz}b. The curves show the THz pulse trace (red), the negligible response to IR light above the ordering temperature at 300K (black), and the response of the spin state which occurs below the Neel temperature (blue). When the crystal is magnetically ordered, the THz response shows a much more dramatic response compared to above room temperature. This is reminiscent of other effects that seem to be enhanced with THz radiation, such as the surprising sensitivity to the superconducting condensate in the presence of charge rather than magnetic order, in the high-temperature superconductor YBCO \cite{dakovski-2015-prb}. These results point to the value of exploiting the use of the strong THz response to magnetic systems with different types of ordering while using the high repetition rate capability at X-FELs to map out the ultrafast magnetic response in fresh detail.

\subsection{X-ray Photon Fluctuation Spectroscopy}
\label{sec:xpfs}
Another area of anticipation is in using x-ray pulses with different separation times between pulses, to perform `probe-probe' measurements to study magnetic fluctuations, sometimes referred to as X-ray Photon Fluctuation Spectroscopy' (XPFS) \cite{Shen-mrsa-2021}. This is similar to X-ray intensity fluctuation spectroscopy (XIFS) \cite{MarkSutton-2008} or X-ray photon correlation spectroscopy (XPCS) \cite{sinha-advmat-2014}, but instead of correlating scattered x-ray speckle patterns between shots, the shots are added together and the contrast is extracted from the pulse pair, when the pulses within each pair are finely spaced. These methods take advantage of the high degree of coherence of the x-rays and advanced light sources to produce `speckle' patterns, where scattered photons create a complex pattern based on the exact configuration of the system, the configuration which is typically not detected but averaged over when the degree of coherence is not high. By monitoring this speckle pattern, fluctuations of the structure can be measured and related back to theory to directly access the interactions between constituents within the system.

XPFS is an ultrafast version of XIFS or XPCS and is in the spirit of speckle visibility spectroscopy, where the contrast is analyzed rather than the intensity-intensity autocorrelation function, but the contrast is obtained from a pair of summed pulses~\cite{gutt2009measuring}. This benefit provides technical motivation because in this case, the area detector collecting images does not have to be read out at the rate of the pulse separation, but rather the repetition rate of the pulse pairs. This capability provides an incredible potential for access to much shorter times. Importantly, the information can still be captured with the summation of the individual pulses, when keeping track of the number of pulses added and with the caveat that the signal-to-noise is lower for multi-shot images.

\begin{figure}
    \centering
    \includegraphics[width=\linewidth]{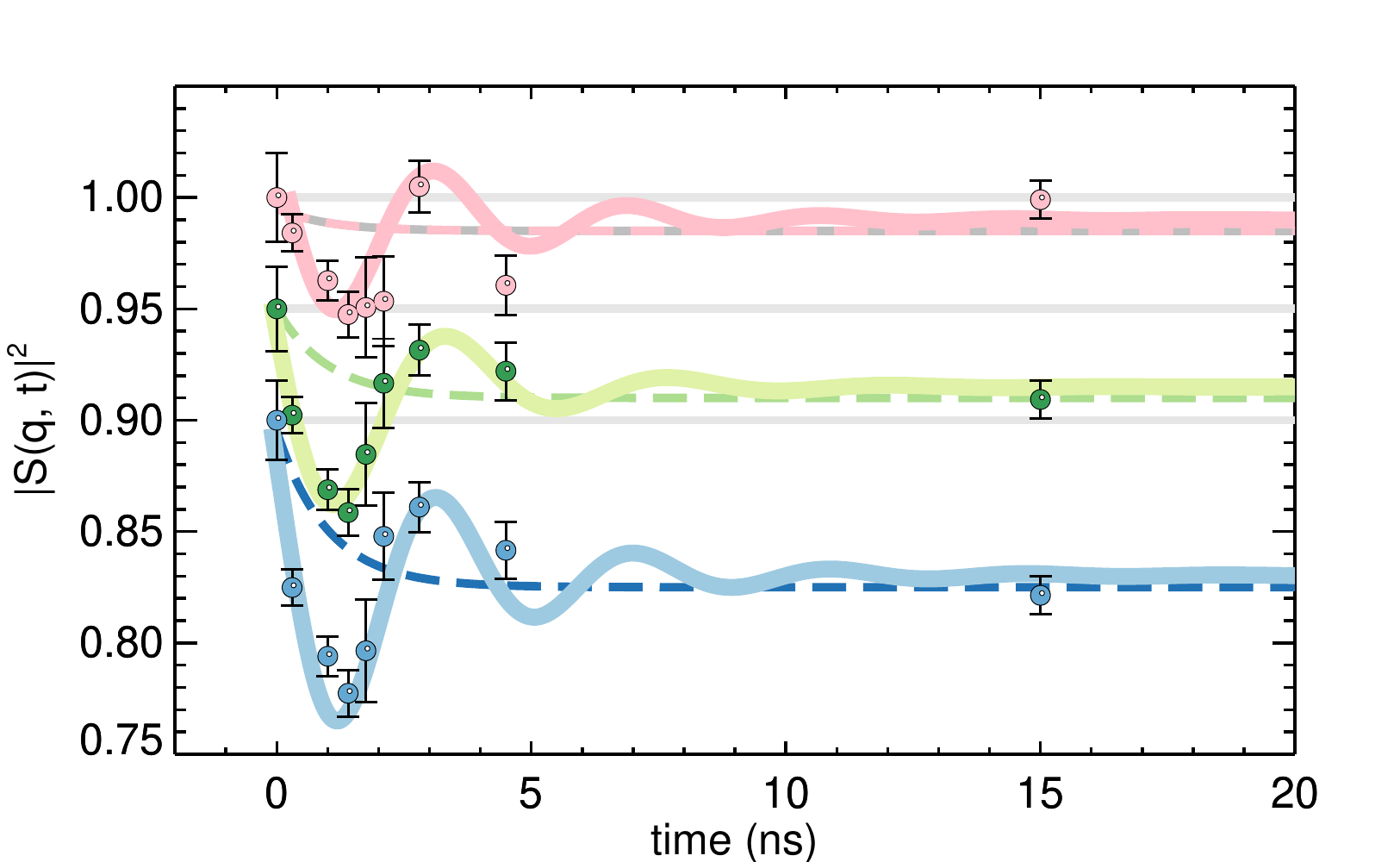}
    \caption{
The time dependent structure factor $S(\mathbf{q}, t)$ of the skyrmion lattice of FeGd recorded at the skyrmion lattice scattering peak. This is a multi-layered system which forms a lattice and spontaneous dynamics can be measured with XPFS on nanosecond times. The correlation function as a function of amplitude ratio $\Lambda = 20\%$ (pink), $12.5\%$ (green), and $10\%$ (blue). The curves are plotted along with an pure exponential decay of the same time constant (dotted lines) are offset by 0.05 for clarity.
    }
    \label{fig:sky}
\end{figure}

By using XPFS, ground state fluctuations can be studied of a magnetic texture in its natural state, without inducing the non-equilibrium nature which is typically associated with the ultrafast. This also requires ultra-short pulses to take snap shots of the magnetic state for different times, but the focus is to make a comparison of the system for different times. For instance, this has been carried out in the multilayered system FeGd \cite{seaberg-2017-prl, esposito-2020-apl,seaberg-2021-prr}, which has been shown to form a skyrmion lattice \cite{Lee-2016-APL,montoya-2017-prb, montoya-2017b-prb}. Here by varying the pulse separation, a damped `phonon-like' mode was measured of the skyrmion lattice. This is reminiscent of the Goldstone mode, and was observed by directly probing the resonant scattering from the magnetic quasi-particles.

One point to note is that, in using this method, equal pulses are needed to extract the fluctuations in the system being studied. We demonstrate this by showing new data from the damped oscillatory fluctuations from the skyrmion lattice mentioned above, in the FeGd system in Fig.~\ref{fig:sky}. In a Self-Amplified Spontaneous Emission (SASE) process such as that produced at the LCLS, each pulse amplitude generated in the accelerator can vary dramatically. This can be captured in the nanosecond regime by using a fast  digitizer as a diagnostic \cite{seaberg-2017-prl,sun-jsr-2018,esposito-2020-apl,assefa-rsi-2022}. In Fig.~\ref{fig:sky}, we analyze pulses that have the pulse ratio between the first and second pulses $\Lambda_{12}=a_1/a_2 - 1 \le 10\%$, or vice versa $\Lambda_{21}=a_2/a_1 - 1 \le 10\%$, shown in the blue curve, as reported by Seaberg et al. \cite{seaberg-2021-prr}. This was determined to be the closest in amplitude one could analyze, while leaving reasonable statistics for this data set. In addition, we show plots with $\Lambda \le 12.5\%$ (green), and $\Lambda \le 20\%$ (pink). These plots demonstrate both the washing out of the oscillation amplitude in the correlation function as well as the larger background at long times. Both of these effects are expected as a result of adding pulse pairs with more `single-pulse' characteristics. For instance, in the extreme case, where one pulse is large while the other is negligible, the contrast will approach $C(\mathbf{q},\tau)\sim1$, the value of a single shot -- regardless of the state of the system. This behavior underpins the importance of this type of diagnostic in XPFS and the need to produce equal pulses in this mode. Future work could correct for this amplitude ratio and retain the largely asymmetric data to improve statistics further. 

Moreover, recent work has focused on changing the ratio of the two probes to increase the first well beyond the second pulse by adjusting the pulse energies generated from the machine (Sec.\,\ref{sec:fresh-slice}). The idea of this option is to perform x-ray pump / x-ray probe by selectively x-ray pumping the system to study the response with soft x-rays. This is quite different in nature to XPFS and instead has the goal of exciting the system at very high energies. 

Finally, we point out the natural extension of this, where one combines XPFS with an optical/THz pump pulse to probe non-equilibrium fluctuations in the excited state. Here the idea would instead be to not keep the system in equilibrium, but understand how the dynamical heterogeneity takes place during the excitation process. This is more sophisticated then a typical pump-probe study because much more information about the length-scale is available than in the scattered intensity. In other words, the monitoring of the contrast can be used for going beyond the coherent response to probe disorder and heterogeneity in non-equilibrium systems as well.

\subsection{Ultrafast Diffuse X-ray Scattering}
A third method that will be critical to exploit in the field of ultrafast magnetism is in the study of diffuse magnetic scattering, where weakly or less well defined long-range scattering can be measured. Features that often define the most interesting properties are in regions where materials exhibit well-defined structure and excitations, and can be resolved in reciprocal space and can even be controlled externally (see Sec.~\ref{pumpprobesec}).  However, sometimes considerable information is contained away from the well-defined structural responses of the system, in the shorter range interactions, which disturb the density-density correlations and lead to diffuse scattering near the peaks corresponding to long-range order. For example, as the thermal motion of atoms about their sites generates thermal diffuse scattering that can be used to obtain information about phonons~\cite{trigo2013fourier}, there is analogous information which will shed light on the magnetic structure as well. In this case, scattering from different components can overlap in reciprocal space, but can be unraveled by observing different signatures in the time domain.

\begin{figure}
    \centering
    \includegraphics[width=\linewidth]{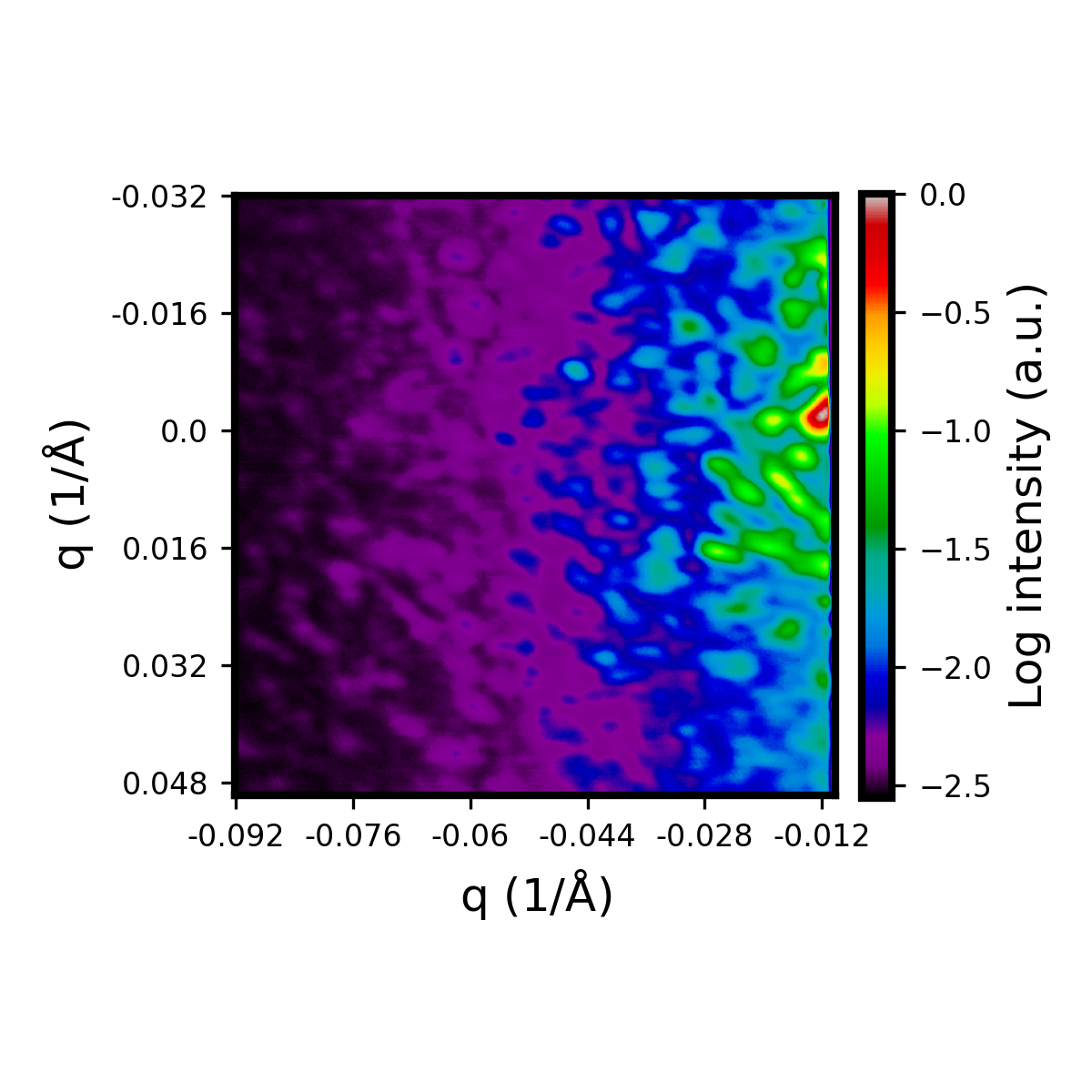}
    \caption{Resonant speckle pattern collected at an X-FEL which can be used to study nanosecond scale spontaneous fluctuations of the prototypical spin-glass system CuMn. The image shows the spin-glass speckle over a 500$\times$500 pixel image using the XPFS prototype instrument \cite{assefa-rsi-2022}. The image was collected at the $L-$edge resonance for Mn in a forward scattering geometry and demonstrates the degree to which magnetism can be extracted with a large amount of diffuse scattering and on short timescales, of order one pulse length of about 100fs.  The center of the peak located $q=0$ is positioned to the right of the sensor region.
    }
    \label{fig:spin-glass}
\end{figure}

For example, in the spin-glass systems, structural or charge scattering can obscure magnetic scattering directly related to the spin structure. In thin films of CuMn, this has been observed by comparing non-resonant to resonant scattering at the Mn-edge as a function of momentum transfer. This was shown to be successful in the measurement of the 4-spin correlation function which displays dynamics on very slow timescales, at the level of hundreds to thousands of seconds, to determine the Edwards-Anderson order parameter \cite{song-2020-arxiv}. Diffuse scattering was furthermore measured in a forward scattering geometry at the LCLS, shown in Fig.~\ref{fig:spin-glass}. This shows a resonant coherent speckle pattern of the diffuse scattering around $q=0$ using the XPFS prototype instrument at the LCLS \cite{assefa-rsi-2022}. It was shown that the large amount of diffuse scattering in the spin glass state could be measured out to large $q$, up to almost 
100\,m{\AA$^{-1}$}, and could furthermore be captured on the order of one pulse width, about 100fs.  Here, instead of using resonant enhancement to separate the charge and magnetic scattering for extraction of pure spin dynamics, the dynamical signatures of the scattering could be used to measure the dynamic spin component, on top of the static charge. 

Another path for magnetic diffuse x-ray scattering is working at a suitable resonant edge, which under the right conditions is advisable to boost the observable magnetic signal. 
To push to the true ultrafast regime requires some of the techniques discussed earlier in this section (see Sec.~\ref{sec:xpfs} for instance), especially with new electron accelerator technology \cite{decker-2022-scirep} (see Sec.~\ref{sec:fresh-slice}). 
A celebrated example of this in magnetism is the observation of the so-called ``pinch points" that are characteristic of the spin ice state \cite{Fennell2007}. Diffuse scattering studies using neutrons for instance have been carried out, where the spin-$\frac{1}{2}$ neutrons interact directly with the magnetic fields generated by magnetic ions.  Such measurements typically take a long time due to the limited neutron flux available, and can usually be considered as long-time averages of the short-range order. As we might expect from the fluctuation-dissipation theorem, this short-range scattering may display dynamics on a range of time scales, depending on the specific origin of the diffuse scattering.  This is well studied in soft condensed matter, for example regarding the glass transition in polymers \cite{hoshino-prl-2020}, but less so in crystalline materials.

On longer time scales, diffuse scattering from neutron scattering on frustrated magnets has been observed to vary on time scales of 10s of minutes, with changes failing to stabilize over several hours in the Ising spin chain Ca$_3$Co$_2$O$_6$ \cite{Agrestini2011}.  In a similar type of system, $\gamma$-CoV$_2$O$_6$, similar changes occur but on a much shorter time scale, with relaxations observed by muon spin rotation decaying over several $\mathrm{\mu}$s \cite{Shen2021}.  Using neutrons, the timescale can be pushed shorter by using the neutron spin-echo technique \cite{Gardner2020}, where timescales ps to 100s of $\mu$s can be probed.  This has also been used extensively in the study of polymers, but again is very flux hungry, with a limited number of instruments around the world.  

Just as with polymer glasses, magnetic glasses, quantum spin glasses, and frustrated magnets are classes of materials where diffuse scattering is vital in being able to unravel the nature of the interactions, which are frustrated either geometrically \cite{Gardner2010} or by introducing site or bond disorder to randomize the exchange interactions \cite{Binder1986}.  Neutron spin echo has been used to study a variety of spin glasses, often showing that significant dynamical processes exist on sub-picosecond timescales \cite{Pappas2006}, but no ultrafast version of any of these neutron scattering methods exists. By targeting the many areas which have been addressed above in magnetic neutron diffuse scattering with ultrafast x-rays, a wide ranging and fertile ground is available for exploration.


\section{Theory}
To take full advantage of the research available with the expanding X-FEL capabilities, certain theoretical tools are vital to tie the methods as outlined above together.  Especially due to the notorious scarcity of X-FEL beamtime and experimental complexity of the methods involved.  Strong predictive models are needed before and during the experiment to ensure the allocated time is used optimally and efficiently.  Furthermore, because of the rich variety of interactions between x-rays and materials, a theoretical perspective is crucial when analyzing data in order to distinguish signal from background. 
Since the focal point here is on resonant x-ray scattering, we start with an overview of the theoretical background for the magnetic cross-section, with an added discussion of  how this relates directly to XPFS. This is followed by a discussion of Density Functional Theory (DFT) and the numerical methods we focus on for magnetism, Exact Diagnolization (ED) and the Density Matrix Renormalization Group (DMRG).

\subsection{Magnetic cross-section and resonant XPFS}
The main mechanism we will focus on in this article is resonant magnetic scattering.  In an experiment this is achieved by tuning the incoming x-ray photon energy to an absorption edge of the metallic ion carrying the magnetic spin-moments originating from unpaired valence electrons.
To model this, we typically focus on dipole transitions, though the quadropolar channel can also be studied.
The full cross-section for resonant scattering for the electric dipole transitions was first worked out by M. Blume \cite{blume-1985-jap} and is given by:
\begin{equation} \label{eq:1}
f = f_c - i f_{m1}(  {\epsilon_{f}^{*}} \times \epsilon_i)\cdot \textbf{s} + {f_{m2}( \epsilon_{f}^{*} \cdot \textbf{s})( \epsilon_i\cdot \textbf{s}) } 
\end{equation}
where $\epsilon$ represents the incoming and final polarization states, $s$ is the spin of the atom, and the $f_i$'s are charge, first, and second order magnetic, frequency-dependent scattering amplitudes.

Typically, in a scattering experiment, we project the cross-section into components that are either in the scattering plane, or orthogonal to the scattering plane. For our purposes discussed in this paper, we tend to focus on the second term, which for small angle scattering gives only off-diagonal matrix elements for the scattering process \cite{hill-1996-actacrys}:

 \[
 {\epsilon_{f}^{*}} \times \epsilon_i=
  \left( \begin{array}{cc}
 {\epsilon_{\sigma}^{*}} \times \epsilon_{\sigma} & {\epsilon_{\sigma}^{*}} \times \epsilon_{\pi}  \\
 {\epsilon_{\pi}^{*}} \times \epsilon_{\sigma} & {\epsilon_{\pi}^{*}} \times \epsilon_{\pi}  \end{array} \right) =
  \left( \begin{array}{cc}
 0 & \bf{k_i}  \\
 \bf{-k_f} & 0
 \end{array} \right)
 \]
In this case, the cross-section goes as $\sim k \cdot s$ and is optimized for spins pointing out of the sample plane, or parallel to the incoming beam.  The resonant intensity enhancement can be several hundredfold that of the non-resonant magnetic contribution, but is still often small compared to the intensity originating from charge-scattering.  In practice, it is best to choose Bragg reflections which are forbidden by the space-group of the chemical lattice but allowed by the magnetic sublattice, so that the first term of \ref{eq:1} can be ignored and only the magnetic terms remain.

In order to observe the dynamics associated with the magnetic scatterers as described in \ref{sec:xpfs} the resultant scattering intensity from \ref{eq:1} is measured as a time-series at a region of interest in reciprocal space $q$. For a typical XPCS experiment, the intensity-intensity autocorrelation function can be calculated:
\begin{equation}
g_2(q,\tau)=\frac{\left<I(q,t)I(q,{t+\tau})\right>}{\left<I(q,t)\right>^2}
\end{equation}
where $\tau$ is the time difference between intensities at different times, and the brackets designate an average over $t$. Importantly, this can be cast in terms of the intermediate scattering function by the Siegert relation \cite{siegert-1943-mitrlr} as:
\begin{equation}
g_2(q,\tau)=1+A{[S(q,\tau)/S(q)]}^2
\end{equation}
where the intermediate scattering function is equal to the field-field correlation function, $g_1(q,\tau)$. This holds as long as the scattering being observed is a Gaussian process, with the phase having an equal probability on the range of $\phi \in [0,2\pi]$~\cite{goodman-2007-book}. One note here is that when representing magnetic x-ray scattering, this is not the spin-spin correlation function, which is typically calculated from theory, but the squared amplitude of the spin-spin correlation function, or a 4-spin correlation function. However, one is also able to carry out so-called `heterodyne' measurements, and so provide access directly to the spin-spin correlation function.

For the XPFS measurements mentioned in Sec.~\ref{sec:xpfs}, one additional element is the relationship of the contrast, which can be directly calculated from the summed pulses, and the correlation function above. An important development in this area was in the demonstration of the equivalence of these two quantities to within a multiplicative factor \cite{Gutt-2009-OptExp}. This indicates that a contrast measurement is able to retrieve the equivalent information about the intermediate scattering function, as in XPCS. 

Lastly, typical experiments rely on both the large pulse energies and the pulse structure of the beam in time, but the scattered photons are usually collected in the sparse limit. In this case, photon counting is necessary to evaluate the contrast. Because the beam is fully coherent, the contrast can be determined by fitting the distribution of photon counts per speckle to the negative binomial distribution, which relates the contrast $C=C(q,\tau)$ to the probability of $k-$events for a given average intensity, $\overline{k}$:

\begin{equation}
    P(k)=A_0(k,M)\left(\frac{{\overline{k}}}{\overline{k} + M}\right)^k\left(\frac{M}{\overline{k} + M}\right)^M,
    \label{eq:nbd}
\end{equation}
where $A_0(k,M)$ is a normalization constant which depends on the contrast and the speckle photon density, given by:

\begin{equation}
A_0(k,M)=\frac{\Gamma(k+M)}{\Gamma(M)\Gamma(k+1)}
\label{eq:gamma}
\end{equation}

In Eq.~\ref{eq:nbd} and Eq.~\ref{eq:gamma}, the dependence is expressed as the number of degrees of freedom of the speckle pattern $M$, or $M=1/\sqrt{C(q,\tau)}$. Since this equation can not be solved analytically, the parameters of the distribution are usually estimated by invoking estimators that are valid in the low $\overline{k}$ limit \cite{sun2020accurate} or by using maximum likelihood estimation \cite{chitturi2022machine}. Under certain conditions, an analytical solution has been shown to exist when multiple $k-$events are observed \cite{Shen-mrsa-2021}.


\subsection{Density Functional Theory}

Next, we outline our theoretical approach, which starts with first-principles density functional theory (DFT) based modeling. The recipe proceeds along the following steps: (1) Advanced density functionals \cite{hohenberg1964inhomogeneous,kohn1965self}, such as the recently constructed SCAN functional \cite{sun2015strongly}, are used to gain a handle on the ground-state electronic, magnetic, and topological structure. Spin-orbit coupling (SOC) effects can be accounted for in these computations \cite{zhang2020competing}. 
(2) First-principles spin-resolved band structures and wavefunctions are then used to evaluate magnetic anisotropy energies, magnetic moments, exchange parameters, anisotropic exchange coefficients, and the Dzyaloshinskii-Moriya interaction (DMI) \cite{moriya1960anisotropic,dzyaloshinsky1958thermodynamic,skyrme1994selected}. 
(3) Informed by first-principles results, material-specific, effective tight-binding model Hamiltonians can be constructed for incorporating effects of electron-electron interactions in our modeling. (4) Using atomistic spin dynamics (ASD) simulations with our model Hamiltonians, we can investigate the evolution of skyrmion states in various materials \cite{muller2019spirit,mentink2010stable}. Finally, (5) the preceding steps can be repeated after revealing effects of strains and magnetic fields on effects such as skyrmion formation. 

The modeling of skyrmion structures within the DFT framework is quite challenging due to the large size of the magnetic unit cells involved. \ Therefore a combination of DFT and ASD simulations are needed to study DMI-induced skyrmions in the presence of an external magnetic field. This has previously been used in CrI$_3$ \cite{behera2019magnetic}. It involves finding the solution to the Landau-Lifshitz-Gilbert (LLG) equation \cite{mentink2010stable}:
 
 \begin{equation}
\frac{dm}{dt}= -|\gamma|m\times\mathcal{H}_{eff}+\alpha(m\times{\frac{dm}{dt}})
\end{equation}
where $\gamma$ is the gyromagnetic ratio, $m$ is the magnetic moment vector for the magnetic atom, $\alpha$ is the Gilbert damping coefficient and $\mathcal{H}_{eff}$ is the effective magnetic field, 

\begin{equation}
\mathcal{\mathcal{H}}_{eff}(x)=-\nabla \mathcal{H}(x), 	
\end{equation}						
where $\mathcal{H}$ is the Hamiltonian of the system, which can be written as:
\begin{equation}
\mathcal{H}=\mathcal{H}_{ex}+\mathcal{H}_{ani}+\mathcal{H}_z+\mathcal{H}_{DMI}						\end{equation}

Here $\mathcal{H}_{ex}$, $\mathcal{H}_{ani}$, $\mathcal{H}_z$, and $\mathcal{H}_{DMI}$ are the exchange, anisotropy, Zeeman, and DMI terms, respectively. This can be expressed more specifically as:

\begin{widetext}
\begin{equation}
\mathcal{H}=-\sum_{<ij>}J_{ij} n_i\cdot n_j - \sum_{ij} K_j (\hat{K}_j \cdot n_i) ^2-\sum_i \mu_i B\cdot n_i-\sum_{<ij>}D_{ij}\cdot (n_i \times n_j) \label{spinhamiltonian}
\end{equation} 
\end{widetext}
where $J_{ij}$ is the Heisenberg symmetric exchange, $K_j$ is the single-ion magnetic anisotropy, $B$ is the external magnetic field, and $D_{ij}$ is the DMI vector. 
Note that $\hat{K}_j$ denotes the direction of the anisotropy and  $m_i=\mu_i \cdot n_i$ is the magnetic moment. 

The first step of this approach is to obtain the electronic structure of a given magnetic material within the DFT framework. We will then use this electronic structure to build a highly accurate real-space low-energy model using Wannier functions as the basis implemented in the Wannier90 code \cite{marzari2012maximally,mostofi2008wannier90}. Once we achieve the low-energy model, we can employ Green’s functions and magnetic force theorem to systematically calculate the exchange parameters of the Heisenberg Hamiltonian following the Korotin approach \cite{korotin2015calculation}. Following this approach, we can get the isotropic exchange parameters, DMI, and the anisotropic exchange between two sites i,and j from the following equations. 
\begin{equation}
 J_{ij}=Im(A_{ij}^{00}-A_{ij}^{xx}-A_{ij}^{yy}-A_{ij}^{zz} ),
\end{equation}
\begin{equation}
J^{ani}=Im(A^{uv}+A^{vu} )
\end{equation}
\begin{equation}
\vec{D}_{ij}^{u} = Re(A_{ij}^{0u}-A_{ij}^{u0}), 
\end{equation}

where $A_{ij}^{uv} = - \frac{1}{\pi}\int_{-\infty}^{E_F}Tr\{\textbf{p}_i^z \textbf{G}_{ij}^u \textbf{p}_j^z \textbf{G}_{ji}^v \}d\epsilon$, ${u,v} = \{0,x,y,z\}$, $\textbf{G}_{ij}$ is the Greens function, and $\textbf{p}_i=\textbf{H}_{ii} (R=0)$ is the intra-atomic part of the Hamiltonian. These magnetic exchange parameters are then used in the ASD simulation to predict the presence of skyrmions. To our satisfaction, this method has been extensively tested for various magnetic materials and predicted their magnetic properties. For example, we have tested this method to obtain exchange parameters for NiPS$_3$ with a zig-zag AFM structure. The first, second, and third nearest neighbor exchange parameters are obtained as 1.2917 meV, 0.1044 meV, and -4.089 meV, respectively, which are similar to the previous data-, 1.093 meV, 0.613 meV, and -3.882 meV, respectively. 

\subsection{Numerical methods}
The heart of this process after using first-principles DFT computations as the starting point input, is in the strongly correlated numerical methods. Studying the ground state of an effective model Hamiltonian derived from DFT computations can uncover the low-energy physics in quantum magnetism. The spin interactions in the Hamiltonian are typically strongly correlated, which requires associated methods, such as exact diagonalization (ED)\cite{weisse2008exact}, variational Monte Carlo\cite{hu2013direct, choo2019two, chen2022systematic},  and density matrix renormalization group (DMRG)\cite{White1992}, to simulate an accurate ground state of super-exchange electrons. 

Since quantum Monte Carlo suffers from the severe ``sign problem”\cite{loh1990sign, troyer2005computational} at low temperatures and the possible system size can be limited in ED\cite{richter2010spin}, we point out that DMRG can have an impact on the ultrafast study of quantum magnets. DMRG is an unbiased numerical method that is widely used for calculating quantum systems. The second generation DMRG \cite{McCulloch_2007,Pirvu_2010,SCHOLLWOCK201196,White2005,Dolfi2012,Stoudenmire2013} is based on the matrix product state (MPS) and matrix product operator (MPO), which for a given Hamiltonian, such as that defined in Eq.~(\ref{spinhamiltonian}), can precisely encode a MPO. This allows the algorithm to target the true ground state by minimizing the variational ground state energy.

\begin{figure}
    \centering
    \includegraphics[width=\linewidth]{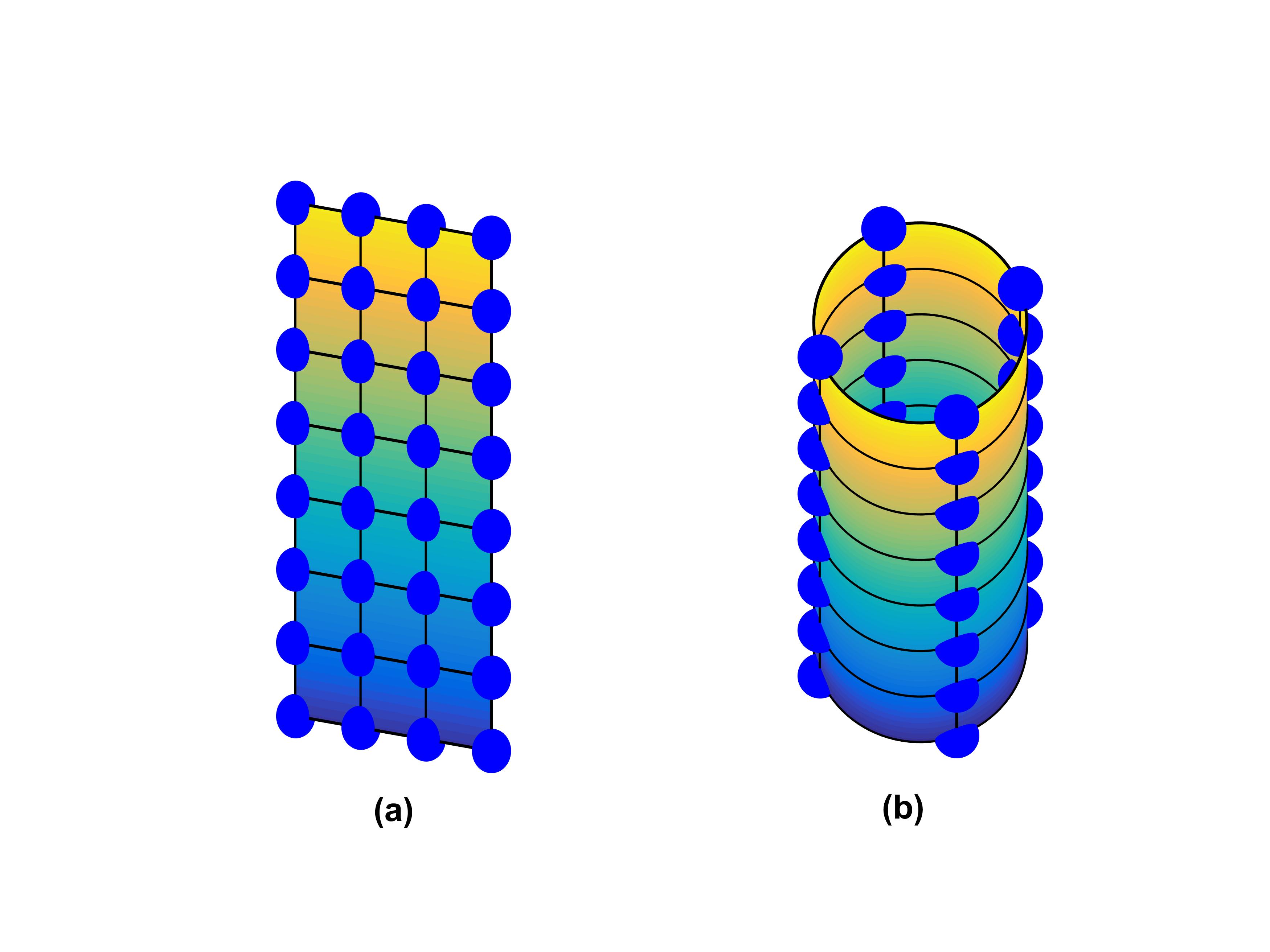}
    \caption{
    The quasi-one-dimensional system is used for the DMRG calculations. (a) An example of a 4-leg ladder, with extension in the vertical direction. (b) For the computation to mimic the 2D condition, a cylinder geometry allows for a periodic boundary condition in one dimension.
    }
    \label{fig:dmrg}
\end{figure}

Searching for this ground state of a given Hamiltonian can be costly due to the exponentially increasing Hilbert space with system size. DMRG is a one-dimensional (1D) algorithm, as initially proposed, but has been used to simulate two-dimensional (2D) systems. This is possible by numbering each site of an interacting spin in real space into a sequence so that some of the nearest neighbor spins in real 2D become long-range pairs in the sequence. The approximation is inevitable in DMRG because it prioritizes short-range entanglement. However, the long-range entanglement, if not dominant, can also be truncated to minimize the computational cost. Therefore, DMRG always utilizes a cylinder geometry (as shown in Fig.\ref{fig:dmrg}) with periodic boundary conditions in one direction
to approximate a 2D system so that the nearest neighbor sites will not be separated by a large distance when transforming to the numbered sequence. The realisitc 2D system is approached by expanding the cylinder width.

Considering conserved quantum numbers can further reduce the computational cost when approaching larger cylinders with higher accuracy. The constraints rule out unwanted states, such as violation of particle number conservation, so that one can search the variational ground state in a subspace faster. The ground state properties of quantum materials can suffer from finite-size effects, low computational accuracy, or both, so it is highly nontrivial to push the limit of large-scale DMRG calculations to achieve higher accuracy with large system sizes. Further speedup is possible through parallel computing.

Moreover, the dynamical density-matrix renormalization group (DDMRG)\cite{White1999} is a variational method for calculating the zero-temperature dynamical properties in 1D and quasi 1D quantum many-body systems. Typically, it is given by a dynamical correlation function, defined as
\begin{equation}
G_{X}(\omega + i \eta)= -\frac{1}{\pi}\langle \psi_0|\hat{X}^{\dag}\frac{1}{E_0+\omega+i\eta-H}\hat{X}|\psi_0\rangle
\end{equation}
Where, $|\psi_0\rangle$ is the ground state and $\hat{X}$ is a quantum operator. However, consider $W_{X,\eta}(\omega,\psi)$ substantively with the formula
\begin{equation}
W_{X,\eta}(\omega,\psi)=\langle \psi|(E_0+\omega -H)^2+\eta^2|\psi\rangle+\eta\langle X|\psi\rangle +\eta\langle \psi|X\rangle
\end{equation}
Where $|X\rangle = \hat{X} |\psi_0\rangle$. Then minimizing $W_{X,\eta}(\omega,\psi)$ yields the imaginary part of the dynamical correlation function for $\eta\rightarrow 0$, {\it i.e.}, $I_X(\omega+ i \eta)= \text{Im } G_{X}(\omega + i \eta)= -W_{X,\eta}(\omega,\psi_{min})/\pi\eta$. The variational problem for solving $|\psi_{min}\rangle$ is achievable with the standard DMRG.

The zero-temperature time-dependent correlation function, defined as $G_{X}(t\geq 0) = \langle \psi_{0}|\hat{X}(t)\hat{X}(0)|\psi_{0}\rangle$, is solvable through the Laplace transform of a spectral function $G_{X}(\omega + i \eta)$, or using ED together with time-dependent DMRG \cite{tdmrg2002}. Comparing the calculated results through the numerical techniques discussed here with real experimental data, can help confirm or further modify the microscopic model provided by DFT. Significantly, this type of modeling of magnetic systems has fresh potential for original contributions in further understanding the results using the experimental X-FEL methods in the ultrafast regime. 

\section{Machine Learning}

In this section, we present a machine learning (ML) model that uses predictive classification to effectively replace the more traditional droplet algorithms that are widely used in XPFS-type measurements. Based on fully convolutional neural networks (CNNs), we show the algorithm is able to photonize raw input XPFS patterns and extract important information under significant levels of electron cloud smearing. We demonstrate that the CNN is able to obtain discrete photon maps from our procedure, which is used to extract the contrast in this method. We previously reported the success of this pipeline using a regression based ML-approach in which the discrete number of photon counts was estimated as a continuous value \cite{chitturi2022machine}. Here, we present an alternative approach in which photon assignments are classified into different discrete bins via optimization with the categorical cross-entropy loss function. 

\subsection{Training Methodology}
\label{sec:training}

To train the model, we simulated a training dataset of 100,000 XPFS patterns with contrasts between 0.7 and 0.9 as well as their corresponding photonized labels. This data was obtained from an  accurate simulator developed for the Linac Coherent Light Source (LCLS) on previous XPFS data, reported in previous work.\cite{burdet2021absolute} Each input pattern has dimensionality 90x90x1 and corresponds to the raw ADU intensity profile of the x-ray measurment. The relevant simulation parameters are the photon energy (340 ADU), the baseline gaussian noise ($N(0, 15)$ ADU) and the probabilistic charge cloud generation (cloud radii $\sigma_g$ values: [0.1,0.25,0.35,0.45,0.55,0.6]). The labels for the machine learning task correspond to reduced images with dimensionality 30x30x1 which contain the true photonized maps. Here, each pixel in the reduced image is matched to the speckle size of the raw data, and stores the number of photons detected per speckle. 

The photonizing task can be formulated as a semantic segmentation computer vision problem \cite{long2015fully}. Specifically, for each pixel in a given XPFS frame we aim to predict a corresponding integer count for the number of photons contained in that pixel. This can be conceptualized as a classification problem where the classes are the photon counts: (0, 1, 2, 3, ..., 8, 8+). To represent each class, we use a one-hot encoding of the class.\cite{goodfellow2016deep} For example, class 3 (representing a 2 photon event) is mapped to [0 0 1 0 0 0 0 0 0]. Therefore, the dimensionality of the output labels changes from 30x30x1 to 30x30x9 in the one-hot encoding representation. We use a U-net neural network model \cite{ronneberger2015u,chitturi2022machine} with the last layer predicting softmax probabilities corresponding to each photon count level and for each pixel $\hat{p}_{i}$ in the resultant photon map, i.e. $\hat{p}_{i,j}$ represents the probability that pixel $i$ is from class $j$. To train the model, we use a cross-entropy loss function ${L}(p_{i}, \hat{p}_{i})$ to measure the difference between the predicted probabilities and true one-hot encoded label ($p_i$), averaged over all pixels in all frames (N).

\begin{equation}
L(p_{i}, \hat{p}_{i}) =-\sum_{i=1}^{\text {N}} p_{i} \cdot \log \hat{p}_{i}
\end{equation}

We obtain an estimate of the contrast by using a per-image estimate for $\bar{k}$ and using the maximum likelihood procedure \cite{roseker2018towards}. Error bars for the contrast follow the asymptotic 95$\%$ CI interval: $\hat{\beta} \pm  \frac{1.96}{\sqrt{n I(\hat{\beta})}}$. The trained neural network model is freely available at \url{https://github.com/src47/CNN_XPFS}.

The ML model discussed here was shown to perform comparably with even the most advanced droplet algorithms, such as the GGG \cite{burdet2021absolute}, on typical simulated XPFS data with a number of evaluation metrics. In addition, since we train our models on simulated data with a known ground truth, we furthermore are able to evaluate the accuracy of the trained CNN, relative to the known absolute contrast value. 

\subsection{Metrics}
\label{sec:metrics}
The metrics used to evaluate the photonizing task are an important part of our evaluation and are described here. For example, the overall accuracy is not a good metric because of data sparsity. In other words, a model which predicts 0 for each pixel will show a uninformatively high accuracy, but this is clearly undesirable. Therefore, we report a set of metric precision scores $\mathcal{P}$, recall $\mathcal{R}$, and the per-class $\mathcal{F}1$ which can be obtained from the number of true positives (\textit{$T_P$}), false positives (\textit{$F_P$}) and false negatives (\textit{$F_N$}), defined as:

\begin{equation}
\label{eqn:metrics_ML}
\begin{split}
&\mathcal{P} = \frac{\textit{\text{$T_P$}}}{\textit{\text{$T_P$}} + \textit{\text{$F_P$}}} \\
&\textit{{$\mathcal{R}$}} = \frac{\textit{\textit{\text{$T_P$}}}}{\textit{\text{$T_P$}} + \textit{\text{$F_N$}}} \\
&\textit{\text{$\mathcal{F}1$ }} = \frac{2 * \textit{\text{$\mathcal{P}$}} * \textit{\text{$\mathcal{R}$}}}{\textit{\text{$\mathcal{P}$}}  + \textit{\text{$\mathcal{R}$}}}
\end{split}
\end{equation}
\\~\\

As an example, the per-class precision for the 2-photon class is the number of true 2-photon events divided by the sum of the true number of 2-photon events and the number of 2-photon events that are incorrectly predicted. In Table \ref{tab:contrast_0.7}, we show the results for three different metrics to evaluate the contrast based on an average contrast of 0.7. The droplet algorithm is compared to the CNN using all three of these metrics, and the results are shown for different photon number events. For this work, we used an average simulated contrast value of between 0.7 and 0.9, as this is what was measured in a previous experiment \cite{seaberg-2017-prl} and was used as a basis for the modeling \cite{burdet2021absolute}. In Table \ref{tab:contrast_0.9}, the same data are displayed for an average simulated contrast value of 0.9.

\begin{figure}[t]
    \centering
        \centering
        \includegraphics[width=0.3\textwidth]{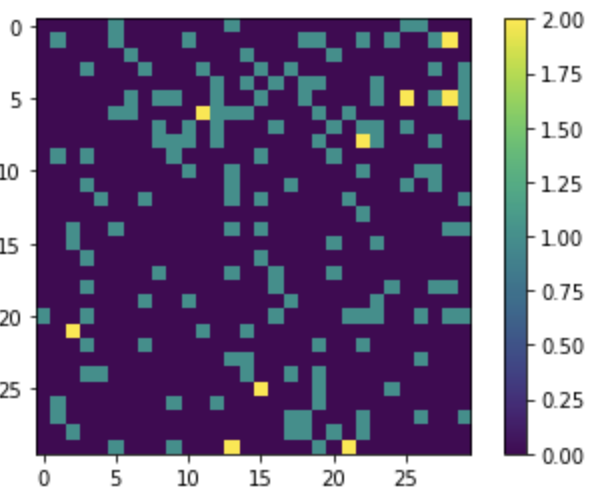}
        \centering
        \includegraphics[width=0.3\textwidth]{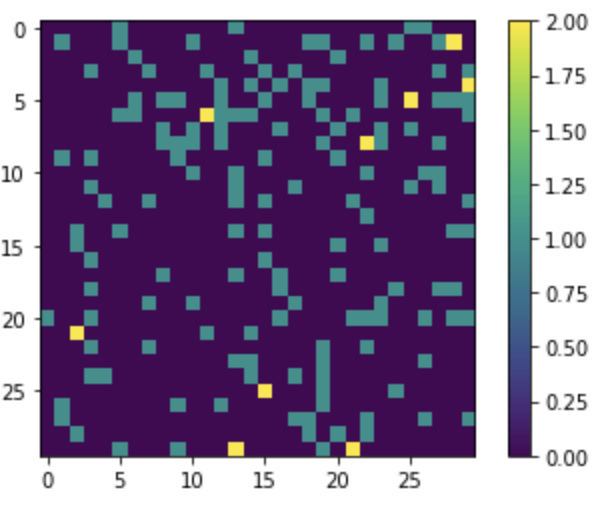}
        \centering
        \includegraphics[width=0.3\textwidth]{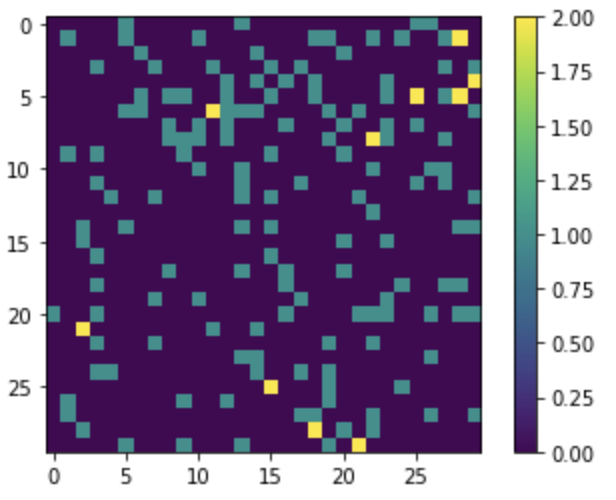}
    \caption{\label{fig:compare_maps}Example of a representative Ground Truth photon map (a), CNN photon map (b) and Droplet photon map (c). On close inspection, it is clear the CNN gets more of the minor features of the photon map predicted correctly.}
\end{figure}

\begin{table*}[t]
\caption{\label{tab:contrast_0.7} Per-class Precision, Recall and F1-score for the CNN and Droplet methods (contrast = 0.7).}
\begin{ruledtabular}
\begin{tabular}{lllll}
Photon Count\footnote{Performance on 5, 6, 7, 8, 8+ photon events is not shown as there is insufficient testing data for these classes.} & Precision (Droplet, CNN) & Recall (Droplet, CNN) &  F1-score (Droplet, CNN) & Data Volume \\
\hline
0 & 1.00, 1.00 & 1.00, 1.00 & 1.00, 1.00 & 3950406 \\
1  &  0.97, 0.97   & 0.96, 0.97   &  0.96 , 0.97    & 488364 \\
2    &   0.90, 0.91   &  0.90, 0.93  &   0.90, 0.92  &  54530 \\
3   &    0.81, 0.83   &   0.83, 0.87  &    0.82, 0.85   &   5992 \\
4   &    0.73, 0.74   &   0.74, 0.77  &    0.74, 0.74   &    637 \\
\end{tabular}
\end{ruledtabular}
\caption{\label{tab:contrast_0.9} Per-class Precision, Recall and F1-score for the CNN and Droplet methods (contrast = 0.9).}

\begin{ruledtabular}
\begin{tabular}{lllll}
Photon Count\footnote{Performance on 5, 6, 7, 8, 8+ photon events is not shown as there is insufficient testing data for these classes.} & Precision (Droplet, CNN) & Recall (Droplet, CNN) &  F1-score (Droplet, CNN) & Data Volume \\
\hline
0 & 1.00, 1.00 & 1.00, 1.00 & 1.00, 1.00 & 3957196 \\
1  &  0.96, 0.97   & 0.96, 0.97   &  0.96 , 0.97    & 472173 \\
2    &   0.91, 0.92   &  0.90, 0.93  &   0.90, 0.92  &  60729 \\
3   &    0.84, 0.86   &   0.83, 0.87  &    0.84, 0.87   &   8536 \\
4   &    0.75, 0.73   &   0.75, 0.77  &    0.75, 0.75   &    1140 \\
\end{tabular}
\end{ruledtabular}
\end{table*}

In all test cases here, the results of the ML algorithm based on classification due to the discrete number of photon events was shown to work just as well as the droplet algorithm. In addition to being much faster, a critical benefit is that there is no user input needed for this algorithm, as is the case for the droplet algorithm, as the model has fully learned how to interpret sparse speckle patterns. 

\subsection{Single photon detection}

\begin{figure}[h]
\includegraphics[width=0.45\textwidth]{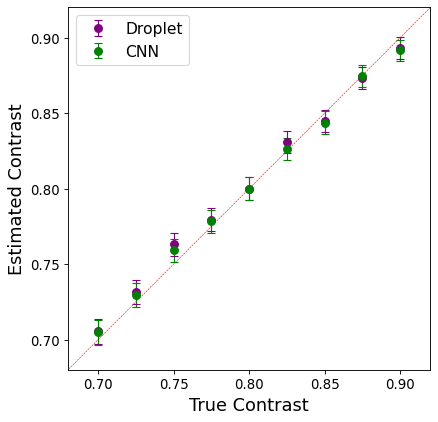}
\caption{\label{fig:contrast_plot} Estimated contrast (the GGG droplet algorithm and CNN) versus the true contrast for both types of photon maps. The value of contrast was chosen to mimic previous experimental contrast value from previous XPFS experiments.}
\end{figure}

A representative example of a true label photon map, the corresponding CNN, and the GGG droplet algorithm predictions are shown in Figure \ref{fig:compare_maps}. Here, it is clear that the CNN and the droplet algorithm agree for most pixels and give visually similar profiles. Although these two methods give similar profiles, small differences in photon count predictions can lead to very different contrast estimates. This is especially true in cases where the counts of high photon events are incorrectly estimated. In Figure \ref{fig:contrast_plot}, we compare contrast estimates derived from both the optimized GGG droplet algorithm\cite{burdet2021absolute} and our CNN model (Figure \ref{fig:contrast_plot}). For the ranges which match the experimental data (contrast values from 0.7 - 0.9), we find that the CNN and the droplet models agree well (Figure \ref{fig:contrast_plot}). 

In order to determine which algorithm gives superior performance in the single photon detection regime, we  evaluate the full performance using the metrics as outlined in Equation \ref{eqn:metrics_ML} for photon maps obtained from both the droplet algorithm and the CNN for the same contrast levels of 0.7 and 0.9. The results are summarized in Table \ref{tab:contrast_0.7}-\ref{tab:contrast_0.9}. In conclusion, the CNN performs comparably, or outperforms, the droplet algorithm in 97\% of the cases across the various metrics. 

\begin{figure}[h]
\includegraphics[width=\linewidth]{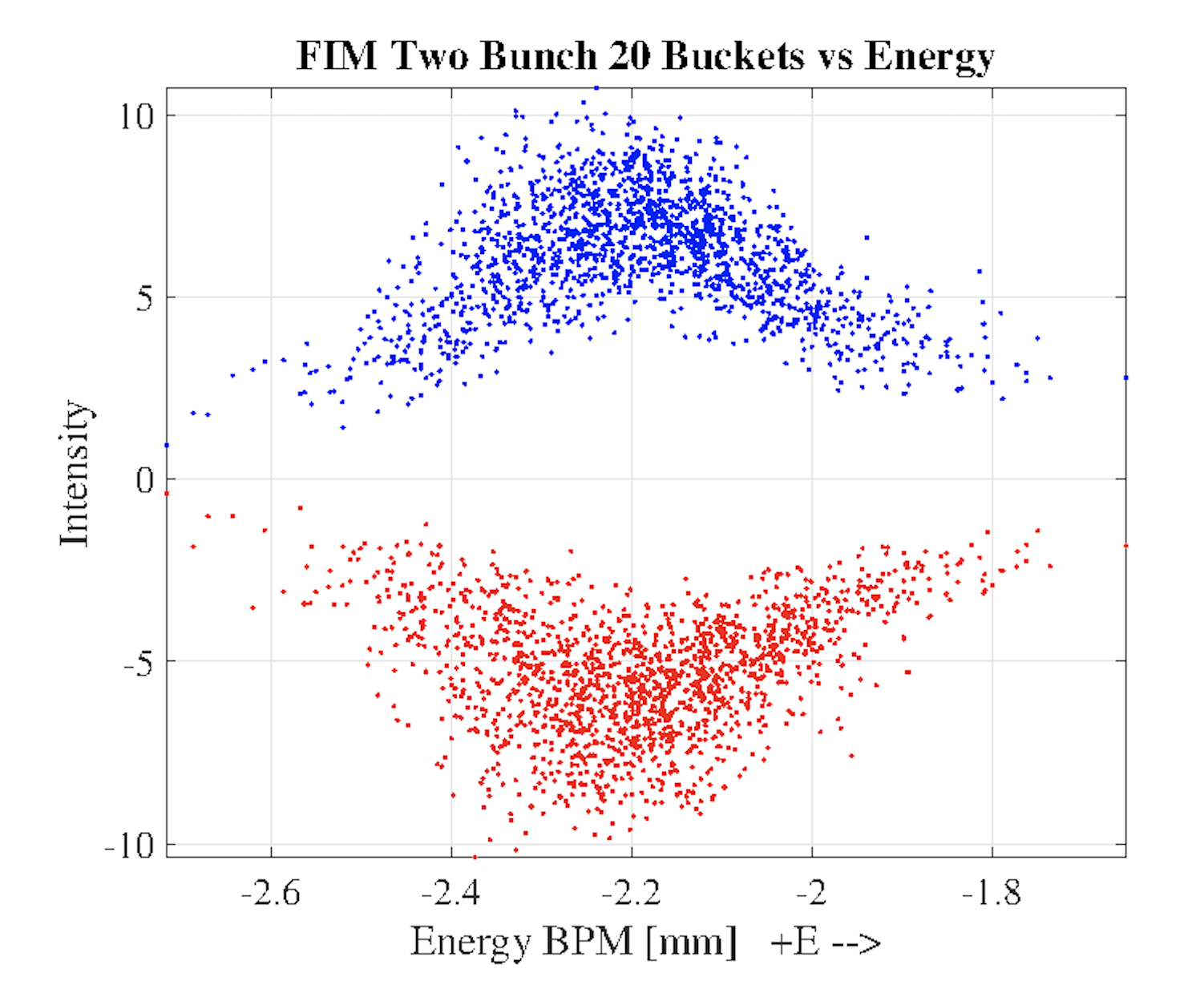}
    \caption{
    {{Two bucket correlation.}}
    The pulse energy of each pulse in a two-bucket approach. The red is plotted negative for convenience. The two peaks are aligned at the same energy, showing they have the same color and can arbitrarily be delayed by integer bucket values of 350ps.
    }
    \label{fig:2bunch}
\end{figure}

As briefly mentioned in Section~\ref{sec:metrics}, it is likely that the CNN model outperforms the GGG algorithm because it it able to learn a variety of filters from labeled datasets constructed using physical knowledge of the scattered photon radiation at the detector. Furthermore, the increased representational power of neural networks allows for the possibility of learning different filters which can handle experimental subtleties, such as variable charge cloud sizes or a wide range of average photon energies \cite{chitturi2022machine}. In contrast, the droplet algorithm is unsupervised and requires substantial tuning `by-hand' to produce optimal results. The results outlined here based on a new ML algorithm to handle XPFS data will be able to accelerate progress in this field and also points to the potential for bridging other efforts in this area, such as the continuous photon modeling approach utilized previously \cite{chitturi2022machine}. Moreover, this also serves to bring more advanced X-ray methods such as this to a wider range of scientific areas, easing the obstacles for general practitioners to execute complex experiments with user-friendly computational tools.

\begin{figure*}
    \centering
     \includegraphics[width=.9\linewidth]{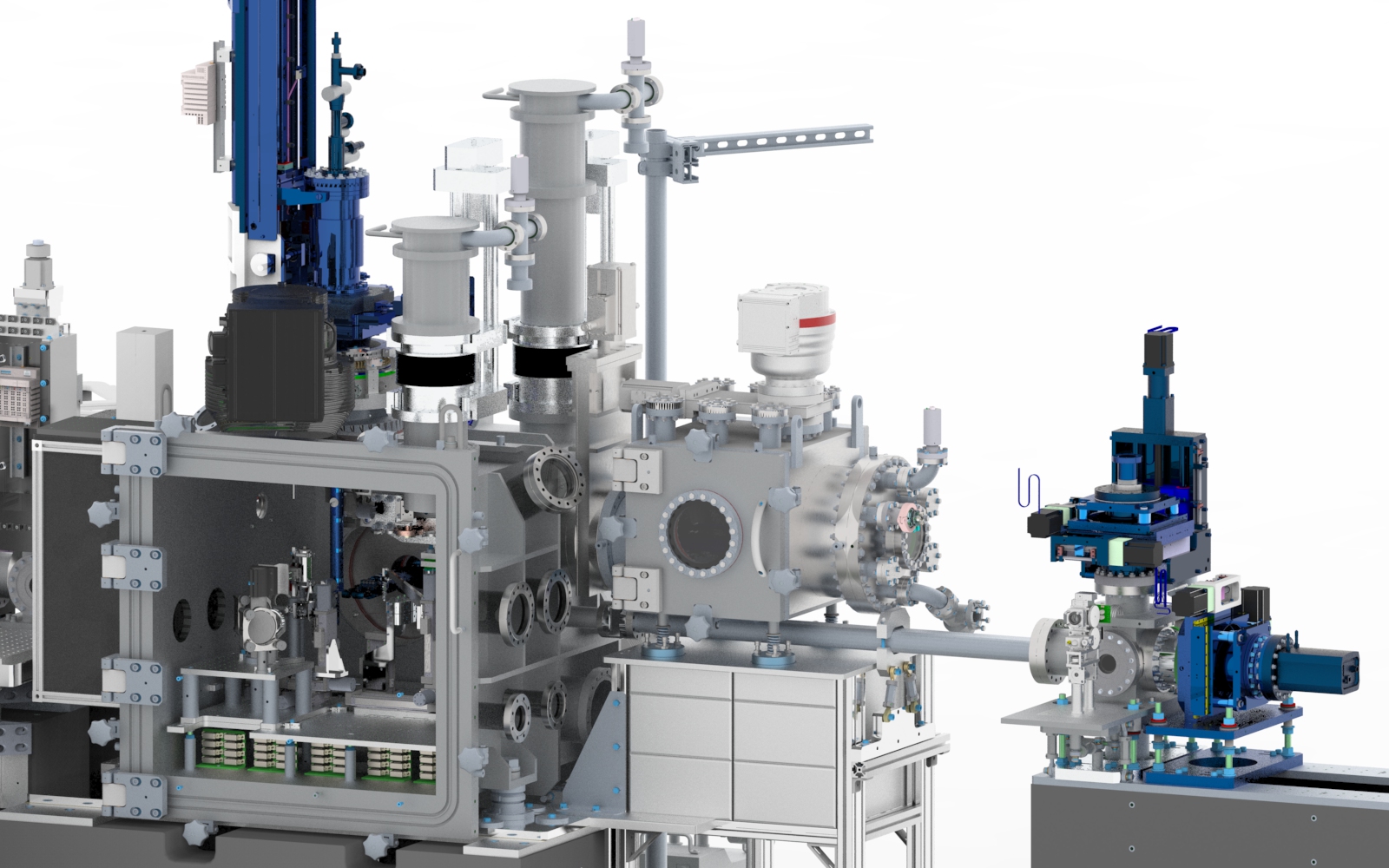}
    \caption{
    {{The chemRIXS Instrument for Solids.}} An image showing the chemRIXS endstation with the enhancements designed for XPFS measurements of solids. The main sample chamber (left) houses the liquid jet system or solid material sample. The electromagnet is inserted perpendicular to the beam horizontally, while the detector and mask assembly (right) are placed in the forward scattering direction for transmission geometry measurements. All dark blue color-coding designate components that were designed for the thin film, ultrafast magnetism experiments.
    }
    \label{fig:chemRIXS}
\end{figure*}

\begin{figure}[h]
    \centering
    \includegraphics[width=.9\linewidth]{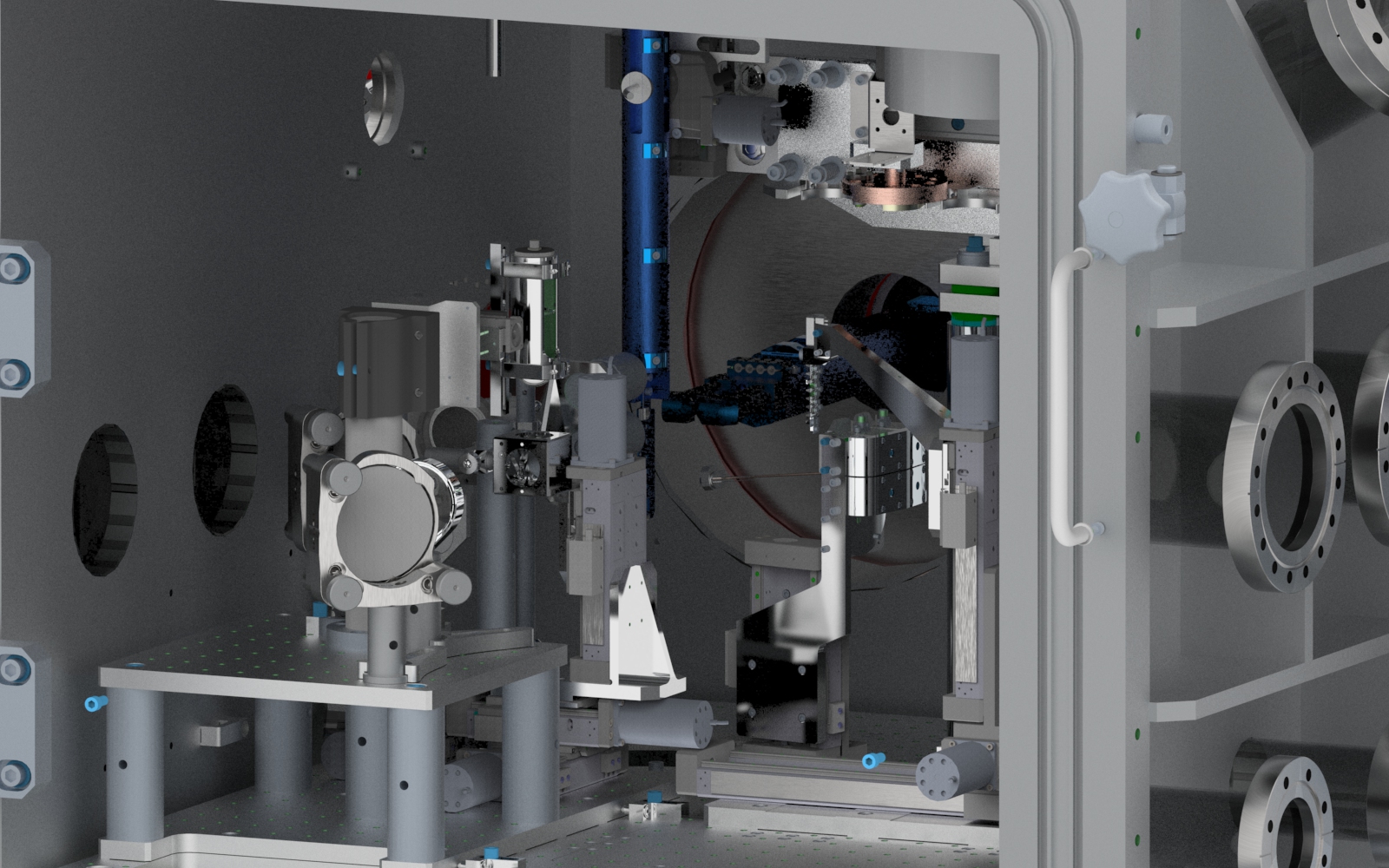}
    \caption{
    {{ChemRIXS Instrument Close-up}} A close-up image inside the sample chamber. This chamber was modified to house an electromagnet and cryostat sample holder to perform XPFS studies on solids.
    }
    \label{fig:chemRIXScloseup}
\end{figure}

\section{Outlook}
\subsection{Scientific Prospects}
With the ultrafast experimental methods outlined here, the prospects for new routes towards understanding magnetism are numerous. THz excitation enables a variety of states to be excited and probed directly, especially with resonant x-ray scattering. By new developments with the pulse structure, x-ray scattering pulses can be compared to study spontaneous fluctuations in the natural ground state of magnetic systems. Finally, diffuse scattering on ultra-fast timescales can also offer new features for the field of magnetism by extending the studies to short range ordered structures, especially in systems that host a large degree of magnetic frustration, or those with short correlation lengths, such as in the spin-glass field. To maximize the potential of these techniques, we discuss in the following special modes of ultra-fast x-ray experiments. We conclude the future outlook with a description about novel instrumentation, and current instruments presently being constructed.

\subsection{Special Accelerator Modes}
\label{sec:accmodes}

Generating different x-ray pulse separation times for performing probe-probe such as XPFS, or x-ray pump / x-ray probe studies as mentioned earlier, is important for executing the ultrafast methods outlined in this paper. This can be accomplished by either optics or special pulse modes developed at the accelerator. Though a large effort has been put into the use of x-ray optics for this task and research progress is tremendous \cite{sun-2019-oe}, here we outline the two most valuable modes in the soft x-ray regime which will be important for the scientific outlook in ultrafast magnetism.

\subsubsection{Fresh-slice x-ray pairs}
\label{sec:fresh-slice}
The Fresh-slice scheme \cite{lutman-2016-natphot} can produce pairs of high-power femtosecond x-ray pulses, with simple control of their wavelength and delay. 
The scheme is based on controlling which temporal slice of an electron bunch lases in a section of the undulator line. This is typically achieved by tilting the electron bunch and subsequently controlling the bunch trajectory. 
Lasing slices can be a couple of femtosecond short, and wavelengths are controlled by the undulator strength in each section, enabling color separation ranging from being at the same wavelength to larger than a factor 2. The maximum delay depends on the strength of an intra-undulator line magnetic chicane. Practically, the scheme has been demonstrated for delays up to 1 picosecond. Temporal coincidence, or exchanging the arrival time of the two pulses, is possible if the slice on the tail is set to lase in the upstream undulator section.

\subsubsection{Two-bunch x-ray pairs}
For longer delays, ranging from hundreds of picoseconds to hundreds of nanoseconds, two separate electron bunches are extracted at the cathode, accelerated and compressed in the linear accelerator and lase in the undulator line \cite{decker-2022-scirep}. The unitary time separation depends on the accelerator RF frequency; for the LCLS S-Band linac, it is close to 350 picoseconds. Performance for each bunch can be similar to that of a regular, single bunch SASE pulse. Typically, the two-bucket scheme is set up to have both pulses lasing at the same wavelength, but a small color separation is achievable by having the two electron bunches at slightly different energies. For instance, Fig.\,\ref{fig:2bunch} shows the pulse energy of each pulse in a dispersive location of the accelerator, with a positive correlation. This can be used to adjust the color of each pulse to have the same wavelength as well as equal pulse energies.

\begin{figure*}
    \includegraphics[width=.99\linewidth]{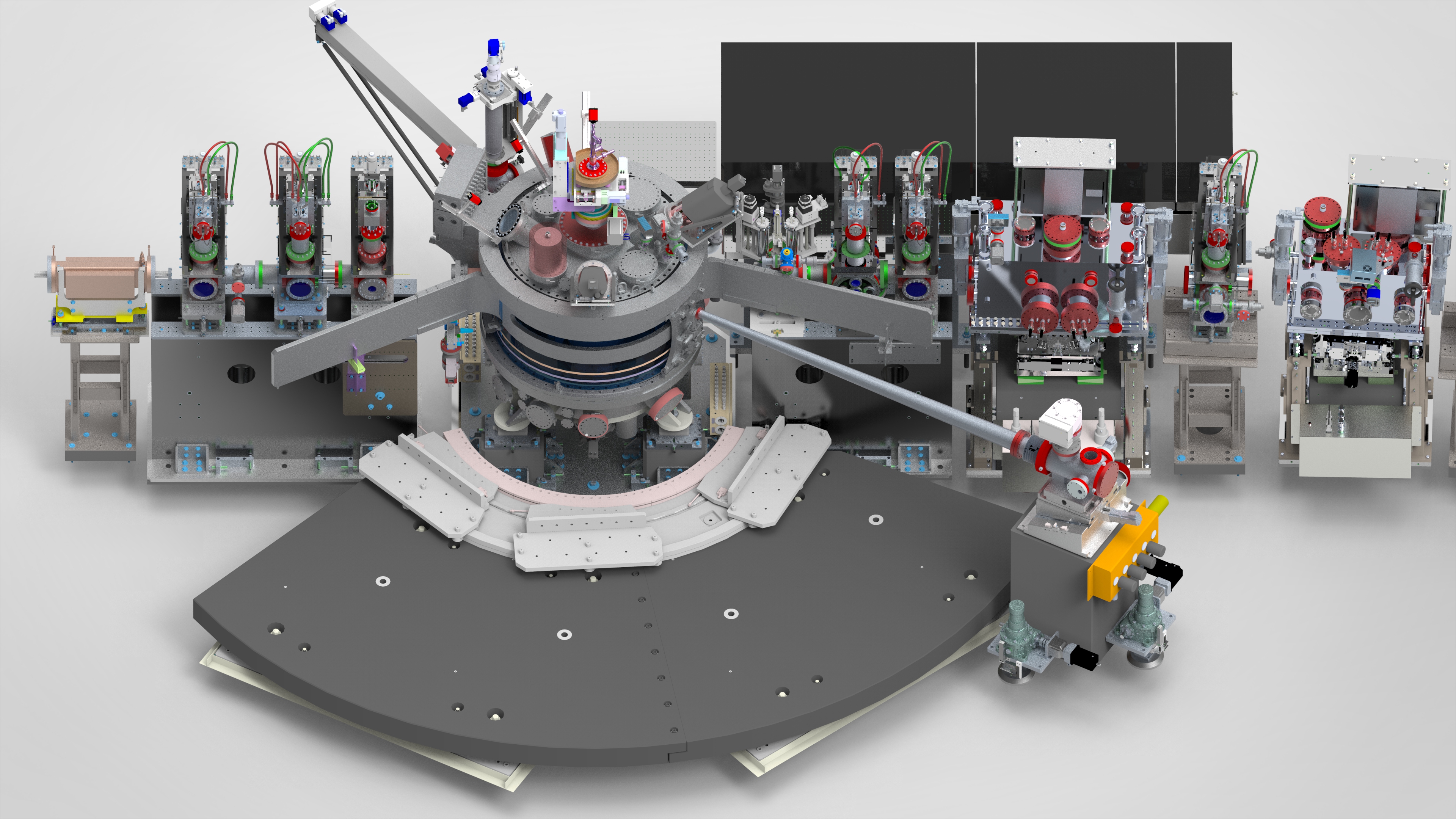}
    \caption{
    {{XPFS Instrument.}} The image shows a rendition of the XPFS chamber housing the fast, pixelated detector, on its own movable stand. The scattering path is evacuated at UHV conditions after precision alignment. The whole apparatus mounts to the qRIXS sample chamber, and is part of the NEH 2.2 instrument at the LCLS-II, shown here installed on the beamline. The instrument is still currently under design.
    }
    \label{fig:chamber}
\end{figure*}

\subsection{State-of-the-art Instrumentation}
Currently, instruments are being developed to take full advantage of the capabilities at these types of sources in the area of magnetism. For instance, at the SLAC National Accelerator Laboratory, the new accelerator for the LCLS-II will soon provide up to nearly $\sim$1\,MHz repetition rate with an array of possible pumping schemes, including THz, with new instruments currently being designed, constructed, and commissioned. Some efforts have been made to retrofit current x-ray instruments, as well as the development of full-scale instruments designed for the purpose of carrying out XPFS measurements in the soft x-ray regime for a variety of different geometries. We outline two such cases here. The first is the modification of the so-called chemRIXS station recently constructed to carry out XPFS studies on materials, while the second is a future instrument currently under construction.

\subsubsection{The chemRIXS instrument for materials}
\label{sec:enh-inst}
As a first test case, we have developed a capability to take advantage of the recent beamline developed as part of the LCLS-II suite of instruments focused on ultrafast chemical science, specifically in x-ray absorption spectroscopy and resonant inelastic x-ray scattering, the so-called ``chemRIXS" beamline. This is specifically focused on using liquid jet or liquid sheet jet systems for the study of ultrafast physics and chemistry experiments \cite{Kunnus-2012-RSI}. The spectrum of the molecules illuminated in the solvent can be optically excited and the chemical structure evolution can be followed on the time-scale of these chemical changes \cite{Wernet-2015-Nature}.

While not designed for solid material samples, some instrumentation enhancements were made such that first XPFS experiments could be performed on the newly designed beamline. This was carried out on thin film magnets by designing and fabricating a solid sample holder to replace the liquid jet, a mounted CCD detector, together with a detector mask, and a manipulator to control the placement of an electromagnet. The mask was used to limit the area of illumination on the detector chip to enable higher speed readout \cite{seaberg-2017-prl}. This feature can be adjusted, i.e.\,to collect a larger fraction of the speckle pattern and a slower rate, depending on the experimental system. The detector was placed in a forward scattering geometry, compatible with the chemRIXS set-up. This detector scheme implemented for this set of experiments has been described in detail elsewhere as a prototype XPFS instrument \cite{assefa-rsi-2022}, but was here used in conjunction with the liquid jet endstation. An overview of this set-up in this configuration is shown in Fig.\,\ref{fig:chemRIXS}.
For a close-up view inside the system, a rendition of the chemRIXS setup from inside the chamber, emphasizing the cryostat and the electromagnetic for magnetic field dependent studies, is shown in Fig.\,\ref{fig:chemRIXScloseup}.

\subsubsection{Future capabilities}
\label{sec:fut-inst}
Looking forward to future capabilities, LCLS-II beamlines are already in use with new instruments on the horizon. For THz pump/ magnetic scattering probe studies or ultrafast magnetic diffuse x-ray scattering for instance, the latest capability will occur with the installation of the qRIXS instrument.

The qRIXS instrument will host a sample chamber and a rotatable spectrometer consisting of grating and high-speed 2D detector covering the range of scattering angles from 40-150 degrees in the horizontal scattering geometry. The spectrometer is designed to achieve an energy resolution of $\sim$ 20 meV at 1 keV. The sample chamber is designed to support elastic soft x-ray diffraction. The chamber is equipped with an in-vacuum diffractometer with a 6-axis degree of freedom of motion.  Sample cooling down to about 25\,K will be possible with the cryogenic sample installation.  The ability to introduce over-sized optics for long-wavelength THz beams, will be accommodated in the near future \cite{Kubacka-2014-Science, Turner-2015-JSR, Hoffmann-2012-SRN}.

In addition the current plan, we are developing another capability to be able to carry out XPFS measurements in the soft x-ray regime as well (see Fig.\ref{fig:chamber}). The focal point of this instrument will be the in-vacuum, long distance, and movable detector motion. This will incorporate the `ePixM' series designed at Stanford University and SLAC National Laboratory. This will be a large, pixelated detector that can count single photons down to the carbon edge ($\sim$ 285 eV) and will operate at the full repetition rate of the new, superconducting accelerator, at 929 kHz. The instrument, which is still under design currently, will be capable of moving the detector about a limited scattering angle range while in UHV conditions, with the use of precision laser trackers. It will be housed in a small vacuum chamber and stand with air bearings for chamber motion. An advanced laser tracker system will place the detector coordinates in the correct geometry at a sample-to-detector distance of about 3\,m. Once aligned, a scattering path length drift tube will be attached from the sample chamber, using a rotary seal, to the detector chamber. It will be evacuated to UHV conditions to accommodate soft x-ray scattering in the area of quantum materials. Once complete, this will offer the potential for first XPFS measurements with soft x-rays at a arbitrary scattering angle, important for a myriad of quantum materials and other types of studies on solids.

\section{Conclusion}
In conclusion, with the latest developments at XFEL sources, studies in ultrafast magnetism are ripe for a renewed focus across many areas of research. We have targeted three new areas that are continued to be developed and will help to capitalize on fresh capabilities to spur magnetic studies in both equilibrium and non-equilibrium investigations. Combining these advanced methods with the latest progress in condensed matter theory and machine learning, the synergy and progress capable in the future for ultrafast magnetism is bright.

\section{Acknowledgements}
This work was supported by the U.S. Department of
Energy, Office of Science, Basic Energy Sciences under Award No.
DE-SC0022216.
 Portions of this work were also supported by the U. S. Department of Energy, Office of Science, Basic Energy Sciences, Materials Sciences and Engineering Division, under Contract DE-AC02-76SF00515. The use of the Linac Coherent Light Source (LCLS), SLAC National Accelerator Laboratory, is also supported by the DOE, Office of Science under the same contract. S. A. M. acknowledges support by the U. S. Office of Naval Research, In-House Lab Independent Research program. S.K., P.F., and S.R. acknowledge support by the U.S. Department of Energy, Office of Science, Office of Basic Energy Sciences, Materials Sciences and Engineering Division under Contract No. DE-AC02-05-CH11231 (NEMM program MSMAG). The research at UCSD was supported by the research programs of the U.S. Department of Energy (DOE), Office of Basic Energy Sciences (Award No. DE-SC0003678).
J. J. Turner acknowledges support from the U.S. DOE, Office of Science, Basic Energy Sciences through the Early Career Research Program.

\bigskip
\bigskip

\bibliography{refs}

\appendix
\renewcommand\thefigure{\thesection.\arabic{figure}}    
\newpage

\end{document}